\documentclass[
 amsmath,
 amssymb,
 aps, 
 prl,
reprint,
]{revtex4-2}

\usepackage{graphicx}
\usepackage{dcolumn}
\usepackage{bm}
\usepackage{hyperref}
\usepackage{tikz-cd}
\usepackage[%
    subpreambles=true,
    sort=true,
    print=true,
    mode=buildnew]{standalone} 
\usepackage{braket} 

\usepackage{xstring,xifthen}

\def\mythresh{1}

\newcommand{\eps}{\varepsilon}
\newcommand{\rev}[1]{%
    \StrLen{#1}[\strlength]
    \ifthenelse{\strlength > \mythresh}{\widetilde{#1}}{\tilde{#1}}
}

\newcommand{\smax}{\Delta s_{\text{max}}}
\begin{document}


\title{Thermodynamic bounds and error correction for faulty coarse graining} 

\author{Jann van der Meer} 
\email{vandermeer.jann.5t@kyoto-u.ac.jp}
\author{Keiji Saito}
\email{keiji.saitoh@scphys.kyoto-u.ac.jp}
\affiliation{%
Department of Physics No. 1, Graduate School of Science, Kyoto University, Kyoto 606-8502, Japan
}

\date{\today}

\begin{abstract}
At the nanoscale, random effects govern not only the dynamics of a physical system but may also affect its observation. This work introduces a novel paradigm for coarse graining that eschews the assignment of a unique coarse-grained trajectory to a microscopic one. Instead, observations are not only coarse-grained but are also accompanied by a small chance of error. Formulating the problem in terms of path weights, we identify a condition on the structure of errors that ensures that the observed entropy production does not increase. As a result, the framework of stochastic thermodynamics for estimating entropy production can be extended to this broader class of systems. As an application, we consider Markov networks in which individual transitions can be observed but may be mistaken for each other. We motivate, derive, and illustrate thermodynamic bounds that relate the error sensitivity of the observed entropy production to the strength of the driving and are valid for arbitrary network topologies. If sufficiently many transitions in the network can be observed, redundancies in the coarse-grained trajectories can be used to detect and correct errors, which potentially improves naive estimates of entropy production. We conclude with an outlook on subsequent research on thermodynamic bounds for erroneous coarse graining.
\end{abstract}

\maketitle


\paragraph{Introduction---} 

The principles of stochastic thermodynamics offer a framework to study energetics and dissipation of microscopic physical systems described by stochastic dynamics \cite{seki10, peli21, shir23, seif25}. Over the past decade, the importance of partially hidden dynamics has been increasingly recognized, and research has expanded beyond deriving exact relations in fully accessible nonequilibrium systems to actively studying specific observables in situations with incomplete information \cite{seif19}. In such settings, results are often expressed as thermodynamic bounds -- inequalities that associate the observable with a suitable, but not directly accessible, thermodynamic quantifier for nonequilibrium like the total entropy production. In recent years, thermodynamic bounds and estimators for entropy production have been formulated based on different observable phenomena including, e.g., precision of currents \cite{bara15, ging16, horo20}, speed limits \cite{aure11, shir18, ito20, vu23} and correlations \cite{dech21, dech23, ohga23} when focusing on particular observables, as well as waiting times in coarse-grained models \cite{skin21a, vdm22, haru22, vdm23, blom24a}. 

A shared assumption of the previous techniques is that coarse graining is reliable, i.e., the microscopic trajectory is related to the coarse-grained one or to the observable via a deterministic mapping. As a consequence, thermodynamic bounds have not yet systematically explored the concept of ``reliability'' of the coarse-graining operation or the observations, although such a concept has proved crucial in thermodynamic tradeoff relations involving speed and dissipation in, e.g., kinetic proofreading and error correction \cite{muru12, sart15, mall20, boeg22}. The scenario of measurement errors may arise naturally if clean observations of, say, states \cite{degu24b, voll24, igos25} and transitions \cite{hart21a, gode23a} become difficult, or if the accessible part of an experimental system, like a colloidal particle \cite{mehl12} or a molecular motor \cite{arig18, naka21}, includes both clearly measurable states and states that are hidden or appear ambiguous.

In this Letter, we describe a framework for coarse graining that allows for the presence of errors and identify a condition under which irreversibility in the time series relates to entropy production as in the error-free case \cite{kawa07, gome08b, rold10}. As an application of this abstract result, we present a model class that generalizes the recently introduced transition-based coarse graining \cite{vdm22, haru22} by allowing for errors in the registered transitions. In this setting, we obtain bounds on the sensitivity of entropy production to the addition of errors. If the faulty coarse-grained trajectory carries some redundant information, appropriate detection or correction mechanisms can enhance these thermodynamic bounds and estimation of entropy production. We expect that investigating these concepts in other system classes will also prove fruitful, and outline future directions in the conclusion.

\paragraph{A framework for faulty coarse graining---} 

We first introduce our notation in the more familiar setting of estimating entropy production from coarse-grained trajectories. We start from the physical assumption that the entropy production rate $\sigma$ (in units of $k_B = 1$) is related to the statistics of the microscopic trajectories $\gamma$ via the general relation \cite{kawa07, gome08b, rold10}
\begin{equation}
    \sigma = \frac{1}{T} \braket{\ln \frac{\mathcal{P}[\gamma]}{\mathcal{P}[\rev{\gamma}]}} = \frac{1}{T} D_{KL}(\mathcal{P}[\gamma]||\mathcal{P}[\rev{\gamma}])
    \label{eq:sigma_def}
,\end{equation}
where the path weight $\mathcal{P}$ quantifies the probability to realize a particular trajectory $\gamma$ of duration $T$ and $\braket{\cdot}$ denotes taking the average. The time-reversal operation $\gamma \mapsto \widetilde{\gamma}$ is given by reading the trajectory backwards, i.e., $\rev{\gamma}(t) = \gamma(T-t)$ if the dynamics is Markovian and describes even variables. 

If an observer cannot access the full underlying dynamics, it is still possible to obtain a lower bound on entropy production provided that the observer sees coarse-grained trajectories $\Gamma$ that result from a many-to-one mapping $\gamma \mapsto \Gamma$. In this case, one can establish \cite{gome08b, rold10, seif19}
\begin{equation}
    \sigma \geq \hat{\sigma} := \frac{1}{T} \braket{\ln \frac{\mathcal{P}[\Gamma]}{\mathcal{P}[\widetilde{\Gamma}]}}
    \label{eq:sigma_est}
.\end{equation}
This reasoning requires that coarse-grained time reversal $\Gamma \mapsto \widetilde{\Gamma}$ is implicitly defined through the requirement $\gamma \mapsto \Gamma \implies \rev{\gamma} \mapsto \widetilde{\Gamma}$ \cite{seif19, hart21a, vdm22, hart24, bisk24}. 

In this work, we consider analogous questions in the more general scenario where coarse graining not only groups microscopic trajectories together but may also introduce errors. Thus, there is an additional layer of randomness between the already inherently stochastic description of the microscopic trajectory, $\gamma$, and the coarse-grained trajectory, $\Gamma^\eps$, that is actually observed. Coarse graining is no longer defined by a deterministic mapping $\gamma \mapsto \Gamma$, but is instead characterized by the likelihood $\mathcal{P}[\Gamma^\eps|\gamma]$ of observing a possible trajectory $\Gamma^\eps$ given the underlying trajectory $\gamma$. In this work we assume that coarse graining and time reversal satisfy
\begin{equation}
    \mathcal{P}[\Gamma^\epsilon|\gamma] = \mathcal{P}[\widetilde{\Gamma^\epsilon}|\widetilde{\gamma}]
    \label{eq:symm_error}
.\end{equation}
Intuitively, this symmetry condition prevents ``biased'' coarse graining by ensuring that an apparent time-asymmetry in $\Gamma^\epsilon$ after coarse graining stems from the irreversibility of $\gamma$ itself. This heuristic is formalized by combining Eq. \eqref{eq:symm_error} with Eq. \eqref{eq:sigma_def} and the chain rule for Kullback-Leibler divergences, which allows us to express
\begin{equation}
    \sigma -\hat{\sigma}^\eps = \frac{1}{T} \braket{\sum_{\gamma} \mathcal{P}[\gamma|\Gamma^\epsilon] \ln \frac{\mathcal{P}[\gamma|\Gamma^\epsilon] }{\mathcal{P}[\rev{\gamma}|\rev{\Gamma^\epsilon}]}}
    \label{eq:sigma_diff}
\end{equation}
when defining $\hat{\sigma}^\eps := \braket{\ln (\mathcal{P}[\Gamma^\eps]/\mathcal{P}[\rev{\Gamma^\eps}])}\!/T$ in close analogy to \eqref{eq:sigma_est}. Since a nonnegative quantity is averaged over $\Gamma^\eps$ in Eq. \eqref{eq:sigma_diff}, we also verify $\sigma \geq \hat{\sigma}^\eps$. We now apply this information-theoretic result to a concrete model class, first considering the case without hidden variables and addressing coarse graining with errors later.

\begin{figure*}[t]
    \raggedright
    \includestandalone[scale=1.0]{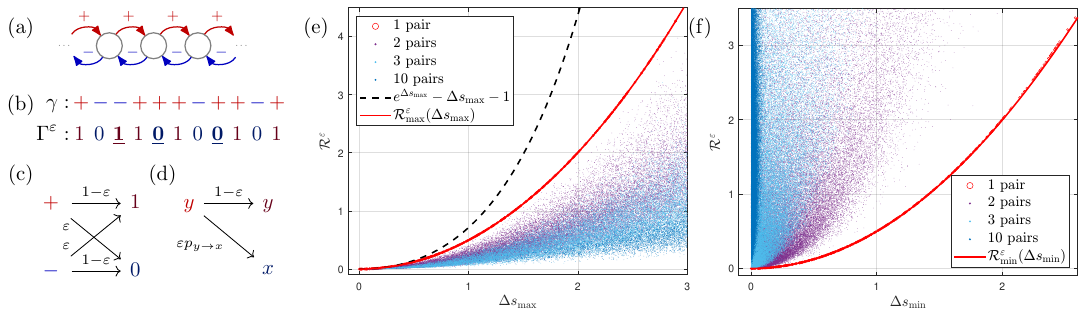}
    \caption{From faulty coarse graining to thermodynamic bounds. (a) On a fundamental level, we assume discrete Markovian dynamics, such as the illustrated random walk. (b) The trajectory $\gamma$ is characterized by a sequence of forward ($+$) and backward ($-$) transitions, but that an external observer cannot measure these steps reliably. The observer's trajectory $\Gamma^\eps$ comprised of registered forward ($1$) and backward ($0$) steps may thus contain errors. (c) The relation between symbols $0,1$ and $+,-$ if errors appear symmetrically with probability $\eps$ (cf. Eq. \eqref{eq:symm_error}). (d) The scheme can be generalized to arbitrary discrete Markovian dynamics, with probability $\eps p_{y \to x}$ to msitake the transition $y$ as $x$. (e) Erroneous measurements affect the estimated entropy production, quantified by $\mathcal{R}^\eps$, the sensitivity of entropy production to errors (Eq. \eqref{eq:multitr:defR}). An upper bound $\mathcal{R}^\eps \leq \mathcal{R}^\eps_{\text{max}} (\Delta s_\text{max})$ (Eq. \eqref{eq:multitr:upperbound}, red curve) is given in terms of the maximal affinity $\Delta s_\text{max}$ and is saturated for the random walk (c). This bound, as a function of $\Delta s_\text{max}$, is compared to $\mathcal{R}^\eps$ of randomly selected systems with different numbers of transition pairs.  (f) A lower bound $\Delta s_{\text{min}}^2/2 \leq \mathcal{R}^\eps$ is identified as a function of the minimal distance of affinities $\Delta s_\text{min}$ (Eq. \eqref{eq:num:lb}, red curve). Model parameters for the simulation are provided in the SM \cite{Supplement}.}
    \label{fig:bounds}
\end{figure*}

\paragraph{Paradigmatic example---}
We first introduce the setting in the simple, paradigmatic example of a discrete asymmetric random walk in continuous time, illustrated in Figure \ref{fig:bounds}\,(a). Denoting the probability of forward and backward jumps by $p(+)$ and $p(-) = 1 - p(+)$, respectively, the average entropy production rate associated with the asymmetric walk is
\begin{equation}
    \sigma = \frac{1}{\braket{t}} \left( p(+) - p(-) \right) \ln \frac{p(+)}{p(-)}
,\end{equation}
with $1/\!\braket{t}$ denoting the mean rate of jumps per unit time. 

Errors are introduced by assuming an observer who cannot measure the sequence of forward and backward jumps that comprise microscopic trajectory $\gamma$ directly. Instead, a possibly faulty trajectory $\Gamma^\eps$ is registered, which is comprised of symbols $1$ or $0$ that indicate forward and backward steps but have a finite chance $\eps$ of erring, as indicated in the scheme given in Figure \ref{fig:bounds}\,(c). As a consequence, observing the potentially defective trajectories $\Gamma^\eps$ gives us access to the probabilities $p(1)$ and $p(0)=1-p(1)$, from which not the true entropy production rate $\sigma$ but rather
\begin{align}
    \hat{\sigma}^\eps = \frac{1}{\braket{t}} \left( p(1) - p(0) \right) \ln \frac{p(1)}{p(0)}
\end{align}
is inferred. Applying our previous reasoning and Eq. \eqref{eq:sigma_diff}, we can conclude that $\sigma^\eps$ is a lower bound on $\sigma$ if Eq. \eqref{eq:symm_error} is satisfied. In the present example, this property is equivalent to the symmetry $p(1|+) = p(0|-)$. Not imposing this condition results in observations that can appear out of equilibrium even for a process that fundamentally satisfies detailed balance $p(+) = p(-)$.

\paragraph{General set-up---}
The set-up can be generalized to arbitrary networks in the stationary state. Let us denote a transition from state $i$ to $j$ by $x = (ij)$ and its reverse by $\rev{x} = (ji)$. We denote the stationary rate of observed transitions $x$ by $p(x)$. The steady-state entropy production rate then takes the form
\begin{align}
    \sigma = \sum_x p(x) \ln \frac{p(x)}{p(\rev{x})}
    \label{eq:multitr:ep_def}
.\end{align}
In the presence of errors, the stationary distribution of transitions $x$ that is accessible to an observer is denoted by $p^\eps(x)$. We parametrize possible errors in the system via
\begin{equation}
    p^\eps(x) = p(x) (1 - \eps) + \sum_y  \eps p(y) p_{y \to x}
    \label{eq:multitr:ep_def}
,\end{equation}
so that with probability $\eps p_{y \to x}$ a transition $y$ is registered as $x$. The chance for an error in the observation of transition $y$, $\eps (1 - p_{y \to y})$, can vary for each transition. To quantify the sensitivity of the observed dissipation to measurement errors, we compare the true entropy production rate $\sigma$ to the value inferred by an observer,
\begin{align}
    \hat{\sigma}^\eps & = \sum_x p^\eps(x) \ln \frac{p^\eps(x)}{p^\eps(\rev{x})} 
.\end{align}

\paragraph{Thermodynamic bounds on response to errors---}

If the probability of errors $\eps \ll 1$ is small, the sensitivity of the entropy production rate to errors can be characterized by a response quantity
\begin{equation}
     \mathcal{R}^\eps = \lim_{\eps \to 0}  \frac{\sigma - \hat{\sigma}^\eps}{\eps} \geq 0
    \label{eq:multitr:defR}
,\end{equation}
whose positivity follows from Eq. \eqref{eq:sigma_diff}. A direct calculation (see Supplemental Material \cite{Supplement}) reveals
\begin{align}
    \mathcal{R}^\eps = \! \sum_{x,y \neq x} \! p_{y \to x} p(y) \left[e^{s(x) - s(y)} - (s(x) - s(y)) - 1 \right]
    \label{eq:multitr:generalres}
,\end{align}
where $s(x) = \ln [p(x)/p(\rev{x})]$ denotes the entropy production associated with a single transition $x$ and will be referred to as the affinity of transition $x$. 

We gain structural insights into Eq. \eqref{eq:multitr:generalres} by identifying thermodynamic upper and lower bounds on $\mathcal{R}^\eps$,
\begin{equation}
    \mathcal{R}^\eps_{\text{l.r.}} \leq \mathcal{R}^\eps  \leq \mathcal{R}^\eps_{\text{max}} (\Delta s_\text{max})
    \label{eq:multitr:inequalities}
.\end{equation}
The upper bound on $\mathcal{R}^\eps$ is formulated in terms of the maximal difference in affinities, which due to the antisymmetry $s(\rev{x}) = -s(x)$ each appear with either sign,
\begin{equation}
 \smax = \max_{x \neq y} |s(x) - s(y)| = 2 \max_{x} |s(x)|
.\end{equation}
A first, crude estimate of the upper bound can be obtained by using normalization, $\sum_{xy} p(y) p_{y \to x} = 1$, together with positivity of $e^z - z -1$, to obtain the estimate $\mathcal{R}^\eps \leq e^{\Delta s_\text{max}} - (\Delta s_\text{max} + 1)$ from Eq. \eqref{eq:multitr:generalres}. A more careful estimate reveals the stronger bound (see SM \cite{Supplement})
\begin{align}
    \mathcal{R}^\eps_{\text{max}} (\Delta s_\text{max}) = 2 \tanh{\frac{\smax}{4}} \left( \frac{\smax}{2} + \sinh \frac{\smax}{2} \right)
    \label{eq:multitr:upperbound}
,\end{align}
which is tight in the case of the asymmetric random walk. A lower bound can be identified using $e^z \geq 1+z+z^2/2$ in Eq. \eqref{eq:multitr:generalres}, which yields
\begin{align}
    \mathcal{R}^\eps_{\text{l.r.}} = (1/2) \sum_{xy} p_{y \to x} p(y) \left( s(x) - s(y) \right)^2
    \label{eq:multitr:lowerbound}
.\end{align}
This lower bound is stronger than Eq. \eqref{eq:sigma_diff} and can be saturated in the linear response regime, where all terms of third and higher order in the affinities $s(x)$ are neglected.

\paragraph{Linear response in terms of entropy production?---} Heuristically, the dependence on $\sim \left( s(x) - s(y) \right)^2$ in Eq. \eqref{eq:multitr:lowerbound} expresses that a significant difference between $\sigma$ and $\sigma^\eps$ requires errors to induce a notable difference in both the affinities and the currents, which, in linear response, are proportional to the affinities themselves. This behavior differs from the scaling of $\sigma$ itself, which takes the form $\sim s(x)^2$ in linear response, thus for small $\sigma$ the ratio $\mathcal{R}^\eps_{\text{l.r.}}/\sigma$ can diverge. As an archetypal example, consider two pairs of transitions $x,y$ with a ``common but cheap'' transition $y$ ($s(y)=0$) and a ``rare but costly'' transition $x$, so that the system can be brought arbitrarily close to equilibrium for small $p(x)$, $p(\rev{x})$. Although $\sigma$ can be made arbitrarily small, the choice $p_{y \to x} = p_{\rev{y} \to \rev{x}} = 1$ leads to a finite response $\mathcal{R}^\eps$ since a significant number of erroneous costly transitions $y$ is registered. In the limit $\sigma \to 0$, the ratio $\mathcal{R}^\eps\!/\sigma$ can become arbitrarily large, but the bounds \eqref{eq:multitr:inequalities} remain meaningful (see Appendix A).

\paragraph{Illustration---} The toy model of the one-dimensional asymmetric random walk can be solved explicitly. Denoting the affinity by $s=p(+)/p(-)$, a direct calculation reveals $\mathcal{R}^\eps = \mathcal{R}^\eps_{\text{max}}$ for $\Delta s_\text{max} = 2s$ (cf. SM \cite{Supplement}). As illustrated in Figure \ref{fig:bounds}, more involved numerical examples with randomly generated $p(x)$ and $p_{y \to x}$ generally do not achieve equality in the bounds \eqref{eq:multitr:inequalities}. The upper bound $\mathcal{R}^\eps \leq \mathcal{R}^\eps_{\text{max}}$ is illustrated in Figure 1\,(c). We also introduce $\Delta s_\text{min}  = \min_{x\neq y} |s(x) - s(y)|$ to obtain a lower bound on $\mathcal{R}^\eps$,
\begin{align}
     \mathcal{R}^\eps \geq \mathcal{R}^\eps_{\text{l.r.}} \geq \mathcal{R}^\eps_{\text{min}} (\Delta s_\text{min})  = (1/2) \left( \Delta s_{\text{min}} \right)^2
    \label{eq:num:lb}
.\end{align}
This bound depends only on $\Delta s_\text{min}$ and is illustrated in Figure 1\,(d).

\begin{figure*}[t]
    \raggedright
    \includestandalone[scale=1.0]{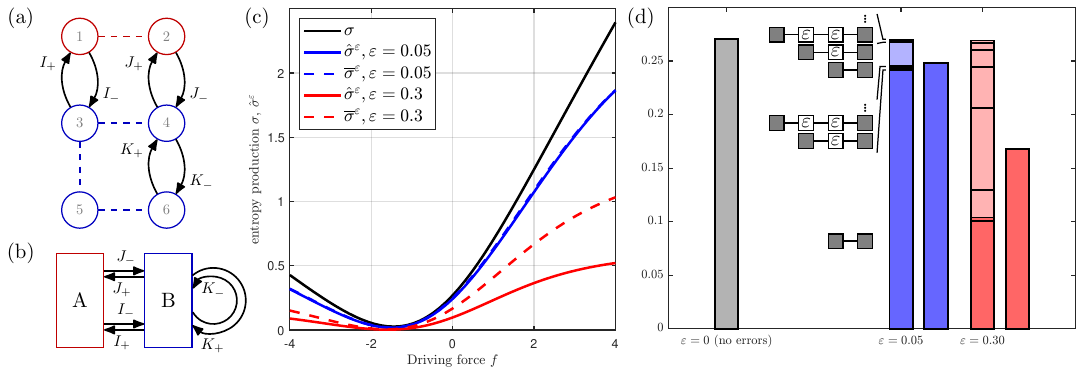}
    \caption{Error correction in a coarse-grained network. (a) Example of a six-state Markov network with three pairs of observed transitions; transition rates are given in the SM \cite{Supplement}. (b) In this topology, the six observed transitions allow to distinguish two connected components $A = \{1, 2 \}$ and $B = \{3, 4, 5, 6\}$. (c) The steady-state entropy production rate $\sigma$ is compared to the estimators $\hat{\sigma}^\eps$ for error rates $\eps=0.05, 0.3$ and varying additional force $f$ along the transition $(34)$. Assuming that errors are isolated, error correction (cf. main text) improves the estimates to $\overline{\sigma}^\eps$ (dashed lines). (d) Decomposition  of $\hat{\sigma}^\eps$ (dark color) and $\sigma - \hat{\sigma}^\eps$ (light color) into contributions $\hat{\sigma}^\eps_n$ and $\sigma_n - \hat{\sigma}^\eps_n$ from sequences with $n=0,1,...$ subsequent errors. The diagrammatic notation show correct transitions as empty gray boxes and erroneous transitions as white boxes containing '$\eps$'. For $\eps=0.05$, contributions due to paths with $n \geq 2$ subsequent errors sum up to $\overline{\sigma}^\eps - \hat{\sigma}^\eps \simeq 5.4 \cdot 10^{-3}$, whereas $\sum_{n \geq 2} \sigma_n \simeq 2.1 \cdot 10^{-3}$. For $\eps=0.3$, we have $\overline{\sigma}^\eps - \hat{\sigma}^\eps \simeq 7.0 \cdot 10^{-2}$ and $\sum_{n \geq 2} \sigma_n \simeq 5.9 \cdot 10^{-2}$. In both cases the improvement due to error correction surpasses the neglected terms of order $\eps^2$ but has a similar order of magnitude.}
    \label{fig:err_corr}
\end{figure*}

\paragraph{Improved bounds through error correction---} Although the previous thermodynamic bounds can be saturated, the results can be improved. The key idea is that the full description of a trajectory often contains redundancies, e.g., if only one pathway connects two states $i,j$ all intermediate states between $i$ and $j$ along that path must have been visited. If some of the redundant information is altered in the observed trajectory $\Gamma^\eps$, we may detect inconsistencies and sometimes even correct them-- a novel feature absent in reliable coarse graining. We illustrate this with a generic example in the case of observed transitions, where $\Gamma^\eps$ emerges from $\gamma$ according to
\begin{alignat}{3}
    \gamma \! : & \, \cdot \to (ij) \to \, \textbf{(jk)} && \to (kl) \to (lm) \to \, \textbf{(mn)} & & \to \cdot \nonumber \\
    \Gamma^\eps \! : & \, \cdot \to (ij) \to \mathbf{(j_\eps k_\eps)} && \to (kl) \to (lm) \to \mathbf{(m_\eps n_\eps)} & & \to \cdot \nonumber
.\end{alignat}
Here, the two transitions $(jk)$ and $(mn)$ are mistaken for $(j_\eps k_\eps) \neq (jk)$ and $(m_\eps n_\eps) \neq (mn)$, respectively. Assuming that subsequent errors are rare and can be neglected, we can detect and correct all errors of the form $y = (jk) \mapsto x = (j_\eps k_\eps)$ that change both the origin and destination ($j \neq j_\eps$ and $k \neq k_\eps$). Effectively, the error probabilities become
\begin{equation}
    \bar{p}_{y \to x} = \begin{cases} 0 & j_\eps \neq j \text{ and } k_\eps \neq k \text{ for } y = (jk) \\ p_{y \to x} & \text{either } j_\eps = j \text{ or } k_\eps = k
    \end{cases}
.\end{equation}
We set $\bar{p}_{y \to y} = 1 - \sum_{x \neq y} \bar{p}_{y \to x}$ to maintain normalization. Combining error correction with the thermodynamic bounds \eqref{eq:multitr:inequalities}, we obtain
\begin{equation}
    \overline{\mathcal{R}}^\eps_{\text{l.r.}} \leq \overline{\mathcal{R}}^\eps  \leq \overline{\mathcal{R}}^\eps_{\text{max}}
    \label{eq:error:inequalities}
\end{equation}
for the error-corrected response $\overline{\mathcal{R}}^\eps$, which takes the form of Eq. \eqref{eq:multitr:generalres} with $p_{y \to x} $ replaced by $\bar{p}_{y \to x}$. Positivity of $f(z) = e^z - z -1$ implies that corrected estimates always improve on non-corrected ones, i.e., $\overline{\mathcal{R}}^\eps \leq \mathcal{R}^\eps$.

\paragraph{Error correction in coarse-grained networks---} We now extend the previous result to the more realistic case that only some transitions in the network can be observed, while others remain completely hidden. Fig. \ref{fig:err_corr}\,(a) shows an example of this transition-based coarse graining \cite{vdm22, haru22} (cf. Appendix B). Assuming for now that transitions can be detected without error, we identify two connected components of the network, $A$ and $B$, shown in Fig. \ref{fig:err_corr}\,(b). Although the microscopic state is not directly accessible, it remains possible to distinguish the connected components $A$ and $B$ based on the latest registered transition. Thus, the six possible transitions can be sorted into four groups $(AA)= \emptyset, (AB) = \{I_-,J_-\}, (BA) = \{I_+,J_+\}, (BB) = \{K_-,K_+\}$ according to their origin and destination. If an error modifies both the origin and destination, it can be detected by the method described above. For example, an error changing $J_+$ into $J_-$ in
\begin{alignat}{10}
    \Gamma \! : & \, \cdot \to K_+ \to \, && \mathbf{J_+} & & \to I_- \to K_+  \to \cdot  \nonumber \\
    \Gamma^\eps \! : & \, \cdot \to K_+ \to \, && \mathbf{J_-} & & \to I_- \to K_+  \to \cdot  \nonumber
\end{alignat}
can be detected, whereas errors changing $J_+$ into $I_+$, a transition of the same group cannot. Modifying either origin or destination, e.g., changing $J_+$ into $K_+$, leads to detectable inconsistencies, but the position of the error cannot be located. Unlike the case where all transitions are observable, we can no longer correct the error with certainty, e.g., a detected error $J_-$ could have originated from either $J_+$ or $I_+$. Nevertheless, a reasonable guess based on knowledge of the system is possible. In this case, we can estimate
\begin{align}
    P(J_+|J_- \text{ detected}) = \frac{p(J_+) p_{J_+ \to J_-}}{\sum_{x = J_+, I_+} p(x) p_{x \to J_-}}
    \label{eq:error:bayes}
\end{align}
using Bayes' theorem provided we know the probabilities that transitions $I_+,J_+$ is registered as $J_-$.

\paragraph{Illustration for finite $\eps$---} The realistic case of a finite error rate is illustrated for the topology of Fig. \ref{fig:err_corr}\,(a). In Fig. \ref{fig:err_corr}\,(c), we compare $\sigma$ to the coarse-grained estimators $\hat{\sigma}^\eps$ and $\overline{\sigma}^\eps$, where the latter incorporates correction of detectable errors as in Eq. \eqref{eq:error:bayes}. Although there is no direct link between trajectory-level error correction and the correction of entropy estimators, we heuristically expect error correction to aid in inferring dissipation because the currents inferred from the error-corrected trajectory are closer to the actual steady-state currents. In our example, the error-corrected estimate is indeed an improvement satisfying $\hat{\sigma}^\eps \leq \overline{\sigma}^\eps \leq \sigma$ for different driving forces $f$ applied at the transition $(34)$ and $\eps=0.05, 0.3$. Note that the error-correction mechanism assumes isolated errors, which may be inappropriate for larger $\eps$ where sequences containing subsequent errors contribute to entropy production substantially. 

Our framework allows to check this hypothesis quantitatively by evaluating Eq. \eqref{eq:sigma_diff}. Since the microscopic dynamics is accessible in this example, we can compute the contributions to $\hat{\sigma}^\eps $ and $\sigma -\hat{\sigma}^\eps$ from individual trajectories $\gamma, \Gamma^\eps$. The assumption of isolated errors is sensible if contributions to $\hat{\sigma}^\eps $ and $\sigma -\hat{\sigma}^\eps$ due to sequences of a trajectory that contain $n \geq 2$ subsequent errors are small. We therefore calculate (see Appendix C for details)
\begin{align}
    \hat{\sigma}^\eps_n = \frac{1}{\braket{t}} \sum_{\gamma} \sum_{\substack{\Gamma^\eps \\ n \text{ errors}}} \mathcal{P}[\Gamma^\eps|\gamma] \mathcal{P}[\gamma]  \ln \frac{\mathcal{P}[\Gamma^\epsilon| \Gamma^\eps_i] }{\mathcal{P}[\rev{\Gamma^\epsilon}|\rev{\Gamma^\eps_f]}}
    \label{eq:error:decomposition}
\end{align}
and, similarly, $\sigma_n$, which collect contributions from steady-state trajectories $\Gamma^\eps$ with $n + 2$ observed transitions: a correct initial transition $\Gamma^\eps_i$, $n$ errors in between, and a correct final transition $\Gamma^\eps_f$. Since any trajectory can be decomposed into such snippets, we have $\hat{\sigma}^\eps = \sum_{n \geq 0} \hat{\sigma}^\eps_n$ and $\sigma = \sum_{n \geq 0} \sigma_n$. Here, $\braket{t}$ denotes the average time between two error-free transitions. The results of this decomposition for the numerical example are displayed in Fig. \ref{fig:err_corr}\,(d). For both $\eps = 0.05$ and $\eps = 0.3$, the improvement in observed dissipation due to error correction is of a similar order of magnitude as the visible and invisible contributions to $\sigma$ from terms with $n \geq 2$, indicating that error-correction is a promising heuristic tool for recovering hidden dissipation even in coarse-grained networks.

\paragraph{Outlook---}

This work explores the impact of errors in coarse graining from the perspective of stochastic thermodynamics. By extending established tools --such as the information-theoretic framework for entropy production-- and introducing novel approaches like error correction, we identify, prove, and illustrate thermodynamic bounds for new model classes that include faulty coarse graining. We anticipate that the relation between dissipation and errors extends beyond the specific case of faulty observations used here as a proof of principle. Thus, our findings may inspire research in other contexts, e.g., misdetected states, continuous observations like waiting times, or continuous state spaces, i.e., underdamped or overdamped Langevin equations, which can be discretized to apply the formalism of this work. Such studies will help distinguish between setting-specific results and general principles, ultimately leading to a broader understanding of the relationship between errors, dissipation and its estimation.

\paragraph{Acknowledgments}
JvdM thanks A. Dechant and K. Kanazawa for stimulating discussions. JvdM was supported by JSPS KAKENHI (Grant No. 24H00833). KS was supported by JSPS KAKENHI (Grant No. JP23K25796).


\begin{thebibliography}{39}%
\makeatletter
\providecommand \@ifxundefined [1]{%
 \@ifx{#1\undefined}
}%
\providecommand \@ifnum [1]{%
 \ifnum #1\expandafter \@firstoftwo
 \else \expandafter \@secondoftwo
 \fi
}%
\providecommand \@ifx [1]{%
 \ifx #1\expandafter \@firstoftwo
 \else \expandafter \@secondoftwo
 \fi
}%
\providecommand \natexlab [1]{#1}%
\providecommand \enquote  [1]{``#1''}%
\providecommand \bibnamefont  [1]{#1}%
\providecommand \bibfnamefont [1]{#1}%
\providecommand \citenamefont [1]{#1}%
\providecommand \href@noop [0]{\@secondoftwo}%
\providecommand \href [0]{\begingroup \@sanitize@url \@href}%
\providecommand \@href[1]{\@@startlink{#1}\@@href}%
\providecommand \@@href[1]{\endgroup#1\@@endlink}%
\providecommand \@sanitize@url [0]{\catcode `\\12\catcode `\$12\catcode `\&12\catcode `\#12\catcode `\^12\catcode `\_12\catcode `\%12\relax}%
\providecommand \@@startlink[1]{}%
\providecommand \@@endlink[0]{}%
\providecommand \url  [0]{\begingroup\@sanitize@url \@url }%
\providecommand \@url [1]{\endgroup\@href {#1}{\urlprefix }}%
\providecommand \urlprefix  [0]{URL }%
\providecommand \Eprint [0]{\href }%
\providecommand \doibase [0]{https://doi.org/}%
\providecommand \selectlanguage [0]{\@gobble}%
\providecommand \bibinfo  [0]{\@secondoftwo}%
\providecommand \bibfield  [0]{\@secondoftwo}%
\providecommand \translation [1]{[#1]}%
\providecommand \BibitemOpen [0]{}%
\providecommand \bibitemStop [0]{}%
\providecommand \bibitemNoStop [0]{.\EOS\space}%
\providecommand \EOS [0]{\spacefactor3000\relax}%
\providecommand \BibitemShut  [1]{\csname bibitem#1\endcsname}%
\let\auto@bib@innerbib\@empty
\bibitem [{\citenamefont {Sekimoto}(2010)}]{seki10}%
  \BibitemOpen
  \bibfield  {author} {\bibinfo {author} {\bibfnamefont {K.}~\bibnamefont {Sekimoto}},\ }\href {https://doi.org/10.1007/978-3-642-05411-2} {\emph {\bibinfo {title} {Stochastic Energetics}}}\ (\bibinfo  {publisher} {Springer},\ \bibinfo {address} {Berlin, Heidelberg},\ \bibinfo {year} {2010})\BibitemShut {NoStop}%
\bibitem [{\citenamefont {Peliti}\ and\ \citenamefont {Pigolotti}(2021)}]{peli21}%
  \BibitemOpen
  \bibfield  {author} {\bibinfo {author} {\bibfnamefont {L.}~\bibnamefont {Peliti}}\ and\ \bibinfo {author} {\bibfnamefont {S.}~\bibnamefont {Pigolotti}},\ }\href@noop {} {\emph {\bibinfo {title} {Stochastic thermodynamics. An Introduction}}}\ (\bibinfo  {publisher} {Princeton Univ. Press},\ \bibinfo {year} {2021})\BibitemShut {NoStop}%
\bibitem [{\citenamefont {Shiraishi}(2023)}]{shir23}%
  \BibitemOpen
  \bibfield  {author} {\bibinfo {author} {\bibfnamefont {N.}~\bibnamefont {Shiraishi}},\ }\href@noop {} {\emph {\bibinfo {title} {An introduction to stochastic thermodynamics: From basic to advanced}}}\ (\bibinfo  {publisher} {Springer Nature Singapore Pte Ltd.},\ \bibinfo {address} {Singapore},\ \bibinfo {year} {2023})\BibitemShut {NoStop}%
\bibitem [{\citenamefont {Seifert}(2025)}]{seif25}%
  \BibitemOpen
  \bibfield  {author} {\bibinfo {author} {\bibfnamefont {U.}~\bibnamefont {Seifert}},\ }\href@noop {} {\emph {\bibinfo {title} {Stochastic thermodynamics}}}\ (\bibinfo  {publisher} {Cambridge University Press},\ \bibinfo {address} {Cambridge, England},\ \bibinfo {year} {2025})\BibitemShut {NoStop}%
\bibitem [{\citenamefont {Seifert}(2019)}]{seif19}%
  \BibitemOpen
  \bibfield  {author} {\bibinfo {author} {\bibfnamefont {U.}~\bibnamefont {Seifert}},\ }\bibfield  {title} {\bibinfo {title} {From stochastic thermodynamics to thermodynamic inference},\ }\href {https://doi.org/10.1146/annurev-conmatphys-031218-013554} {\bibfield  {journal} {\bibinfo  {journal} {Ann. Rev. Cond. Mat. Phys.}\ }\textbf {\bibinfo {volume} {10}},\ \bibinfo {pages} {171} (\bibinfo {year} {2019})}\BibitemShut {NoStop}%
\bibitem [{\citenamefont {Barato}\ and\ \citenamefont {Seifert}(2015)}]{bara15}%
  \BibitemOpen
  \bibfield  {author} {\bibinfo {author} {\bibfnamefont {A.~C.}\ \bibnamefont {Barato}}\ and\ \bibinfo {author} {\bibfnamefont {U.}~\bibnamefont {Seifert}},\ }\bibfield  {title} {\bibinfo {title} {Thermodynamic uncertainty relation for biomolecular processes},\ }\href {https://doi.org/10.1103/PhysRevLett.114.158101} {\bibfield  {journal} {\bibinfo  {journal} {Phys.\ Rev.\ Lett.}\ }\textbf {\bibinfo {volume} {114}},\ \bibinfo {pages} {158101} (\bibinfo {year} {2015})}\BibitemShut {NoStop}%
\bibitem [{\citenamefont {Gingrich}\ \emph {et~al.}(2016)\citenamefont {Gingrich}, \citenamefont {Horowitz}, \citenamefont {Perunov},\ and\ \citenamefont {England}}]{ging16}%
  \BibitemOpen
  \bibfield  {author} {\bibinfo {author} {\bibfnamefont {T.~R.}\ \bibnamefont {Gingrich}}, \bibinfo {author} {\bibfnamefont {J.~M.}\ \bibnamefont {Horowitz}}, \bibinfo {author} {\bibfnamefont {N.}~\bibnamefont {Perunov}},\ and\ \bibinfo {author} {\bibfnamefont {J.~L.}\ \bibnamefont {England}},\ }\bibfield  {title} {\bibinfo {title} {Dissipation bounds all steady-state current fluctuations},\ }\href {https://doi.org/10.1103/PhysRevLett.116.120601} {\bibfield  {journal} {\bibinfo  {journal} {Phys.\ Rev.\ Lett.}\ }\textbf {\bibinfo {volume} {116}},\ \bibinfo {pages} {120601} (\bibinfo {year} {2016})}\BibitemShut {NoStop}%
\bibitem [{\citenamefont {Horowitz}\ and\ \citenamefont {Gingrich}(2020)}]{horo20}%
  \BibitemOpen
  \bibfield  {author} {\bibinfo {author} {\bibfnamefont {J.~M.}\ \bibnamefont {Horowitz}}\ and\ \bibinfo {author} {\bibfnamefont {T.~R.}\ \bibnamefont {Gingrich}},\ }\bibfield  {title} {\bibinfo {title} {Thermodynamic uncertainty relations constrain non-equilibrium fluctuations},\ }\href {https://doi.org/https://doi.org/10.1038/s41567-019-0702-6} {\bibfield  {journal} {\bibinfo  {journal} {Nat. Phys.}\ }\textbf {\bibinfo {volume} {16}},\ \bibinfo {pages} {15} (\bibinfo {year} {2020})}\BibitemShut {NoStop}%
\bibitem [{\citenamefont {Aurell}\ \emph {et~al.}(2011)\citenamefont {Aurell}, \citenamefont {Mej{\'i}a-Monasterio},\ and\ \citenamefont {Muratore-Ginanneschi}}]{aure11}%
  \BibitemOpen
  \bibfield  {author} {\bibinfo {author} {\bibfnamefont {E.}~\bibnamefont {Aurell}}, \bibinfo {author} {\bibfnamefont {C.}~\bibnamefont {Mej{\'i}a-Monasterio}},\ and\ \bibinfo {author} {\bibfnamefont {P.}~\bibnamefont {Muratore-Ginanneschi}},\ }\bibfield  {title} {\bibinfo {title} {Optimal protocols and optimal transport in stochastic thermodynamics},\ }\href {https://doi.org/10.1103/PhysRevLett.106.250601} {\bibfield  {journal} {\bibinfo  {journal} {Phys.\ Rev.\ Lett.}\ }\textbf {\bibinfo {volume} {106}},\ \bibinfo {pages} {250601} (\bibinfo {year} {2011})}\BibitemShut {NoStop}%
\bibitem [{\citenamefont {Shiraishi}\ \emph {et~al.}(2018)\citenamefont {Shiraishi}, \citenamefont {Funo},\ and\ \citenamefont {Saito}}]{shir18}%
  \BibitemOpen
  \bibfield  {author} {\bibinfo {author} {\bibfnamefont {N.}~\bibnamefont {Shiraishi}}, \bibinfo {author} {\bibfnamefont {K.}~\bibnamefont {Funo}},\ and\ \bibinfo {author} {\bibfnamefont {K.}~\bibnamefont {Saito}},\ }\bibfield  {title} {\bibinfo {title} {Speed limit for classical stochastic processes},\ }\href {https://doi.org/10.1103/PhysRevLett.121.070601} {\bibfield  {journal} {\bibinfo  {journal} {Phys. Rev. Lett.}\ }\textbf {\bibinfo {volume} {121}},\ \bibinfo {pages} {070601} (\bibinfo {year} {2018})}\BibitemShut {NoStop}%
\bibitem [{\citenamefont {Ito}\ and\ \citenamefont {Dechant}(2020)}]{ito20}%
  \BibitemOpen
  \bibfield  {author} {\bibinfo {author} {\bibfnamefont {S.}~\bibnamefont {Ito}}\ and\ \bibinfo {author} {\bibfnamefont {A.}~\bibnamefont {Dechant}},\ }\bibfield  {title} {\bibinfo {title} {Stochastic time evolution, information geometry, and the {Cram\'er}-{Rao} bound},\ }\href {https://doi.org/10.1103/PhysRevX.10.021056} {\bibfield  {journal} {\bibinfo  {journal} {Phys. Rev. X}\ }\textbf {\bibinfo {volume} {10}},\ \bibinfo {pages} {021056} (\bibinfo {year} {2020})}\BibitemShut {NoStop}%
\bibitem [{\citenamefont {Van~Vu}\ and\ \citenamefont {Saito}(2023)}]{vu23}%
  \BibitemOpen
  \bibfield  {author} {\bibinfo {author} {\bibfnamefont {T.}~\bibnamefont {Van~Vu}}\ and\ \bibinfo {author} {\bibfnamefont {K.}~\bibnamefont {Saito}},\ }\bibfield  {title} {\bibinfo {title} {Thermodynamic {{Unification}} of {{Optimal Transport}}: {{Thermodynamic Uncertainty Relation}}, {{Minimum Dissipation}}, and {{Thermodynamic Speed Limits}}},\ }\href {https://doi.org/10.1103/PhysRevX.13.011013} {\bibfield  {journal} {\bibinfo  {journal} {Physical Review X}\ }\textbf {\bibinfo {volume} {13}},\ \bibinfo {pages} {011013} (\bibinfo {year} {2023})}\BibitemShut {NoStop}%
\bibitem [{\citenamefont {Dechant}\ and\ \citenamefont {Sasa}(2021)}]{dech21}%
  \BibitemOpen
  \bibfield  {author} {\bibinfo {author} {\bibfnamefont {A.}~\bibnamefont {Dechant}}\ and\ \bibinfo {author} {\bibfnamefont {S.~I.}\ \bibnamefont {Sasa}},\ }\bibfield  {title} {\bibinfo {title} {Improving thermodynamic bounds using correlations},\ }\href {https://doi.org/https://doi.org/10.1103/PhysRevX.11.041061} {\bibfield  {journal} {\bibinfo  {journal} {Phys. Rev. X}\ }\textbf {\bibinfo {volume} {11}},\ \bibinfo {pages} {041061} (\bibinfo {year} {2021})}\BibitemShut {NoStop}%
\bibitem [{\citenamefont {Dechant}\ \emph {et~al.}(2023)\citenamefont {Dechant}, \citenamefont {{Garnier-Brun}},\ and\ \citenamefont {Sasa}}]{dech23}%
  \BibitemOpen
  \bibfield  {author} {\bibinfo {author} {\bibfnamefont {A.}~\bibnamefont {Dechant}}, \bibinfo {author} {\bibfnamefont {J.}~\bibnamefont {{Garnier-Brun}}},\ and\ \bibinfo {author} {\bibfnamefont {S.-i.}\ \bibnamefont {Sasa}},\ }\href@noop {} {\bibinfo {title} {Thermodynamic bounds on correlation times}} (\bibinfo {year} {2023}),\ \Eprint {https://arxiv.org/abs/2303.13038} {arXiv:2303.13038 [cond-mat]} \BibitemShut {NoStop}%
\bibitem [{\citenamefont {Ohga}\ \emph {et~al.}(2023)\citenamefont {Ohga}, \citenamefont {Ito},\ and\ \citenamefont {Kolchinsky}}]{ohga23}%
  \BibitemOpen
  \bibfield  {author} {\bibinfo {author} {\bibfnamefont {N.}~\bibnamefont {Ohga}}, \bibinfo {author} {\bibfnamefont {S.}~\bibnamefont {Ito}},\ and\ \bibinfo {author} {\bibfnamefont {A.}~\bibnamefont {Kolchinsky}},\ }\bibfield  {title} {\bibinfo {title} {Thermodynamic {{Bound}} on the {{Asymmetry}} of {{Cross-Correlations}}},\ }\href {https://doi.org/10.1103/PhysRevLett.131.077101} {\bibfield  {journal} {\bibinfo  {journal} {Physical Review Letters}\ }\textbf {\bibinfo {volume} {131}},\ \bibinfo {pages} {077101} (\bibinfo {year} {2023})}\BibitemShut {NoStop}%
\bibitem [{\citenamefont {Skinner}\ and\ \citenamefont {Dunkel}(2021)}]{skin21a}%
  \BibitemOpen
  \bibfield  {author} {\bibinfo {author} {\bibfnamefont {D.~J.}\ \bibnamefont {Skinner}}\ and\ \bibinfo {author} {\bibfnamefont {J.}~\bibnamefont {Dunkel}},\ }\bibfield  {title} {\bibinfo {title} {Estimating entropy production from waiting time distributions},\ }\href {https://doi.org/https://doi.org/10.1103/PhysRevLett.127.198101} {\bibfield  {journal} {\bibinfo  {journal} {Phys.\ Rev.\ Lett.}\ }\textbf {\bibinfo {volume} {127}},\ \bibinfo {pages} {198101} (\bibinfo {year} {2021})}\BibitemShut {NoStop}%
\bibitem [{\citenamefont {{van der Meer}}\ \emph {et~al.}(2022)\citenamefont {{van der Meer}}, \citenamefont {Ertel},\ and\ \citenamefont {Seifert}}]{vdm22}%
  \BibitemOpen
  \bibfield  {author} {\bibinfo {author} {\bibfnamefont {J.}~\bibnamefont {{van der Meer}}}, \bibinfo {author} {\bibfnamefont {B.}~\bibnamefont {Ertel}},\ and\ \bibinfo {author} {\bibfnamefont {U.}~\bibnamefont {Seifert}},\ }\bibfield  {title} {\bibinfo {title} {Thermodynamic inference in partially accessible markov networks: A unifying perspective from transition-based waiting time distributions},\ }\href {https://doi.org/https://doi.org/10.1103/PhysRevX.12.031025} {\bibfield  {journal} {\bibinfo  {journal} {Phys. Rev. X}\ }\textbf {\bibinfo {volume} {12}},\ \bibinfo {pages} {031025} (\bibinfo {year} {2022})}\BibitemShut {NoStop}%
\bibitem [{\citenamefont {Harunari}\ \emph {et~al.}(2022)\citenamefont {Harunari}, \citenamefont {Dutta}, \citenamefont {Polettini},\ and\ \citenamefont {Roldan}}]{haru22}%
  \BibitemOpen
  \bibfield  {author} {\bibinfo {author} {\bibfnamefont {P.}~\bibnamefont {Harunari}}, \bibinfo {author} {\bibfnamefont {A.}~\bibnamefont {Dutta}}, \bibinfo {author} {\bibfnamefont {M.}~\bibnamefont {Polettini}},\ and\ \bibinfo {author} {\bibfnamefont {E.}~\bibnamefont {Roldan}},\ }\bibfield  {title} {\bibinfo {title} {What to learn from a few visible transitions’ statistics?},\ }\href {https://doi.org/https://doi.org/10.1103/PhysRevX.12.041026} {\bibfield  {journal} {\bibinfo  {journal} {Phys. Rev. X}\ }\textbf {\bibinfo {volume} {12}},\ \bibinfo {pages} {041026} (\bibinfo {year} {2022})}\BibitemShut {NoStop}%
\bibitem [{\citenamefont {Van Der~Meer}\ \emph {et~al.}(2023)\citenamefont {Van Der~Meer}, \citenamefont {Deg{\"u}nther},\ and\ \citenamefont {Seifert}}]{vdm23}%
  \BibitemOpen
  \bibfield  {author} {\bibinfo {author} {\bibfnamefont {J.}~\bibnamefont {Van Der~Meer}}, \bibinfo {author} {\bibfnamefont {J.}~\bibnamefont {Deg{\"u}nther}},\ and\ \bibinfo {author} {\bibfnamefont {U.}~\bibnamefont {Seifert}},\ }\bibfield  {title} {\bibinfo {title} {Time-{{Resolved Statistics}} of {{Snippets}} as {{General Framework}} for {{Model-Free Entropy Estimators}}},\ }\href {https://doi.org/10.1103/PhysRevLett.130.257101} {\bibfield  {journal} {\bibinfo  {journal} {Physical Review Letters}\ }\textbf {\bibinfo {volume} {130}},\ \bibinfo {pages} {257101} (\bibinfo {year} {2023})}\BibitemShut {NoStop}%
\bibitem [{\citenamefont {Blom}\ \emph {et~al.}(2024)\citenamefont {Blom}, \citenamefont {Song}, \citenamefont {Vouga}, \citenamefont {Godec},\ and\ \citenamefont {Makarov}}]{blom24a}%
  \BibitemOpen
  \bibfield  {author} {\bibinfo {author} {\bibfnamefont {K.}~\bibnamefont {Blom}}, \bibinfo {author} {\bibfnamefont {K.}~\bibnamefont {Song}}, \bibinfo {author} {\bibfnamefont {E.}~\bibnamefont {Vouga}}, \bibinfo {author} {\bibfnamefont {A.}~\bibnamefont {Godec}},\ and\ \bibinfo {author} {\bibfnamefont {D.~E.}\ \bibnamefont {Makarov}},\ }\bibfield  {title} {\bibinfo {title} {Milestoning estimators of dissipation in systems observed at a coarse resolution},\ }\href {https://doi.org/10.1073/pnas.2318333121} {\bibfield  {journal} {\bibinfo  {journal} {Proceedings of the National Academy of Sciences}\ }\textbf {\bibinfo {volume} {121}},\ \bibinfo {pages} {e2318333121} (\bibinfo {year} {2024})}\BibitemShut {NoStop}%
\bibitem [{\citenamefont {Murugan}\ \emph {et~al.}(2012)\citenamefont {Murugan}, \citenamefont {Huse},\ and\ \citenamefont {Leibler}}]{muru12}%
  \BibitemOpen
  \bibfield  {author} {\bibinfo {author} {\bibfnamefont {A.}~\bibnamefont {Murugan}}, \bibinfo {author} {\bibfnamefont {D.~A.}\ \bibnamefont {Huse}},\ and\ \bibinfo {author} {\bibfnamefont {S.}~\bibnamefont {Leibler}},\ }\bibfield  {title} {\bibinfo {title} {Speed, dissipation, and error in kinetic proofreading},\ }\href {https://doi.org/10.1073/pnas.1119911109} {\bibfield  {journal} {\bibinfo  {journal} {Proceedings of the National Academy of Sciences}\ }\textbf {\bibinfo {volume} {109}},\ \bibinfo {pages} {12034} (\bibinfo {year} {2012})}\BibitemShut {NoStop}%
\bibitem [{\citenamefont {Sartori}\ and\ \citenamefont {Pigolotti}(2015)}]{sart15}%
  \BibitemOpen
  \bibfield  {author} {\bibinfo {author} {\bibfnamefont {P.}~\bibnamefont {Sartori}}\ and\ \bibinfo {author} {\bibfnamefont {S.}~\bibnamefont {Pigolotti}},\ }\bibfield  {title} {\bibinfo {title} {Thermodynamics of error correction},\ }\href {https://doi.org/10.1103/PhysRevX.5.041039} {\bibfield  {journal} {\bibinfo  {journal} {Phys. Rev. X}\ }\textbf {\bibinfo {volume} {5}},\ \bibinfo {pages} {041039} (\bibinfo {year} {2015})}\BibitemShut {NoStop}%
\bibitem [{\citenamefont {Mallory}\ \emph {et~al.}(2020)\citenamefont {Mallory}, \citenamefont {Igoshin},\ and\ \citenamefont {Kolomeisky}}]{mall20}%
  \BibitemOpen
  \bibfield  {author} {\bibinfo {author} {\bibfnamefont {J.~D.}\ \bibnamefont {Mallory}}, \bibinfo {author} {\bibfnamefont {O.~A.}\ \bibnamefont {Igoshin}},\ and\ \bibinfo {author} {\bibfnamefont {A.~B.}\ \bibnamefont {Kolomeisky}},\ }\bibfield  {title} {\bibinfo {title} {Do {{We Understand}} the {{Mechanisms Used}} by {{Biological Systems}} to {{Correct Their Errors}}?},\ }\href {https://doi.org/10.1021/acs.jpcb.0c06180} {\bibfield  {journal} {\bibinfo  {journal} {The Journal of Physical Chemistry B}\ }\textbf {\bibinfo {volume} {124}},\ \bibinfo {pages} {9289} (\bibinfo {year} {2020})}\BibitemShut {NoStop}%
\bibitem [{\citenamefont {Boeger}(2022)}]{boeg22}%
  \BibitemOpen
  \bibfield  {author} {\bibinfo {author} {\bibfnamefont {H.}~\bibnamefont {Boeger}},\ }\bibfield  {title} {\bibinfo {title} {Kinetic {{Proofreading}}},\ }\href {https://doi.org/10.1146/annurev-biochem-040320-103630} {\bibfield  {journal} {\bibinfo  {journal} {Annual Review of Biochemistry}\ }\textbf {\bibinfo {volume} {91}},\ \bibinfo {pages} {423} (\bibinfo {year} {2022})}\BibitemShut {NoStop}%
\bibitem [{\citenamefont {Deg{\"u}nther}\ \emph {et~al.}(2024)\citenamefont {Deg{\"u}nther}, \citenamefont {{van der Meer}},\ and\ \citenamefont {Seifert}}]{degu24b}%
  \BibitemOpen
  \bibfield  {author} {\bibinfo {author} {\bibfnamefont {J.}~\bibnamefont {Deg{\"u}nther}}, \bibinfo {author} {\bibfnamefont {J.}~\bibnamefont {{van der Meer}}},\ and\ \bibinfo {author} {\bibfnamefont {U.}~\bibnamefont {Seifert}},\ }\bibfield  {title} {\bibinfo {title} {General theory for localizing the where and when of entropy production meets single-molecule experiments},\ }\href {https://doi.org/10.1073/pnas.2405371121} {\bibfield  {journal} {\bibinfo  {journal} {Proceedings of the National Academy of Sciences}\ }\textbf {\bibinfo {volume} {121}},\ \bibinfo {pages} {e2405371121} (\bibinfo {year} {2024})}\BibitemShut {NoStop}%
\bibitem [{\citenamefont {Vollmar}\ \emph {et~al.}(2024)\citenamefont {Vollmar}, \citenamefont {Bebon}, \citenamefont {Schimpf}, \citenamefont {Flietel}, \citenamefont {Celiksoy}, \citenamefont {S{\"o}nnichsen}, \citenamefont {Godec},\ and\ \citenamefont {Hugel}}]{voll24}%
  \BibitemOpen
  \bibfield  {author} {\bibinfo {author} {\bibfnamefont {L.}~\bibnamefont {Vollmar}}, \bibinfo {author} {\bibfnamefont {R.}~\bibnamefont {Bebon}}, \bibinfo {author} {\bibfnamefont {J.}~\bibnamefont {Schimpf}}, \bibinfo {author} {\bibfnamefont {B.}~\bibnamefont {Flietel}}, \bibinfo {author} {\bibfnamefont {S.}~\bibnamefont {Celiksoy}}, \bibinfo {author} {\bibfnamefont {C.}~\bibnamefont {S{\"o}nnichsen}}, \bibinfo {author} {\bibfnamefont {A.}~\bibnamefont {Godec}},\ and\ \bibinfo {author} {\bibfnamefont {T.}~\bibnamefont {Hugel}},\ }\bibfield  {title} {\bibinfo {title} {Model-free inference of memory in conformational dynamics of a multi-domain protein},\ }\href {https://doi.org/10.1088/1751-8121/ad6d1e} {\bibfield  {journal} {\bibinfo  {journal} {Journal of Physics A: Mathematical and Theoretical}\ }\textbf {\bibinfo {volume} {57}},\ \bibinfo {pages} {365001} (\bibinfo {year} {2024})}\BibitemShut {NoStop}%
\bibitem [{\citenamefont {Igoshin}\ \emph {et~al.}(2025)\citenamefont {Igoshin}, \citenamefont {Kolomeisky},\ and\ \citenamefont {Makarov}}]{igos25}%
  \BibitemOpen
  \bibfield  {author} {\bibinfo {author} {\bibfnamefont {O.~A.}\ \bibnamefont {Igoshin}}, \bibinfo {author} {\bibfnamefont {A.~B.}\ \bibnamefont {Kolomeisky}},\ and\ \bibinfo {author} {\bibfnamefont {D.~E.}\ \bibnamefont {Makarov}},\ }\bibfield  {title} {\bibinfo {title} {Uncovering dissipation from coarse observables: {{A}} case study of a random walk with unobserved internal states},\ }\href {https://doi.org/10.1063/5.0247331} {\bibfield  {journal} {\bibinfo  {journal} {The Journal of Chemical Physics}\ }\textbf {\bibinfo {volume} {162}},\ \bibinfo {pages} {034111} (\bibinfo {year} {2025})}\BibitemShut {NoStop}%
\bibitem [{\citenamefont {Hartich}\ and\ \citenamefont {Godec}(2021)}]{hart21a}%
  \BibitemOpen
  \bibfield  {author} {\bibinfo {author} {\bibfnamefont {D.}~\bibnamefont {Hartich}}\ and\ \bibinfo {author} {\bibfnamefont {A.}~\bibnamefont {Godec}},\ }\bibfield  {title} {\bibinfo {title} {Emergent memory and kinetic hysteresis in strongly driven networks},\ }\href {https://doi.org/https://doi.org/10.1103/PhysRevX.11.041047} {\bibfield  {journal} {\bibinfo  {journal} {Phys. Rev. X}\ }\textbf {\bibinfo {volume} {11}},\ \bibinfo {pages} {041047} (\bibinfo {year} {2021})}\BibitemShut {NoStop}%
\bibitem [{\citenamefont {Godec}\ and\ \citenamefont {Makarov}(2023)}]{gode23a}%
  \BibitemOpen
  \bibfield  {author} {\bibinfo {author} {\bibfnamefont {A.}~\bibnamefont {Godec}}\ and\ \bibinfo {author} {\bibfnamefont {D.~E.}\ \bibnamefont {Makarov}},\ }\bibfield  {title} {\bibinfo {title} {Challenges in {{Inferring}} the {{Directionality}} of {{Active Molecular Processes}} from {{Single-Molecule Fluorescence Resonance Energy Transfer Trajectories}}},\ }\href {https://doi.org/10.1021/acs.jpclett.2c03244} {\bibfield  {journal} {\bibinfo  {journal} {The Journal of Physical Chemistry Letters}\ }\textbf {\bibinfo {volume} {14}},\ \bibinfo {pages} {49} (\bibinfo {year} {2023})}\BibitemShut {NoStop}%
\bibitem [{\citenamefont {Mehl}\ \emph {et~al.}(2012)\citenamefont {Mehl}, \citenamefont {Lander}, \citenamefont {Bechinger}, \citenamefont {Blickle},\ and\ \citenamefont {Seifert}}]{mehl12}%
  \BibitemOpen
  \bibfield  {author} {\bibinfo {author} {\bibfnamefont {J.}~\bibnamefont {Mehl}}, \bibinfo {author} {\bibfnamefont {B.}~\bibnamefont {Lander}}, \bibinfo {author} {\bibfnamefont {C.}~\bibnamefont {Bechinger}}, \bibinfo {author} {\bibfnamefont {V.}~\bibnamefont {Blickle}},\ and\ \bibinfo {author} {\bibfnamefont {U.}~\bibnamefont {Seifert}},\ }\bibfield  {title} {\bibinfo {title} {Role of hidden slow degrees of freedom in the fluctuation theorem},\ }\href {https://doi.org/10.1103/PhysRevLett.108.220601} {\bibfield  {journal} {\bibinfo  {journal} {Phys. Rev. Lett.}\ }\textbf {\bibinfo {volume} {108}},\ \bibinfo {pages} {220601} (\bibinfo {year} {2012})}\BibitemShut {NoStop}%
\bibitem [{\citenamefont {Ariga}\ \emph {et~al.}(2018)\citenamefont {Ariga}, \citenamefont {Tomishige},\ and\ \citenamefont {Mizuno}}]{arig18}%
  \BibitemOpen
  \bibfield  {author} {\bibinfo {author} {\bibfnamefont {T.}~\bibnamefont {Ariga}}, \bibinfo {author} {\bibfnamefont {M.}~\bibnamefont {Tomishige}},\ and\ \bibinfo {author} {\bibfnamefont {D.}~\bibnamefont {Mizuno}},\ }\bibfield  {title} {\bibinfo {title} {Nonequilibrium energetics of molecular motor kinesin},\ }\href {https://doi.org/10.1103/PhysRevLett.121.218101} {\bibfield  {journal} {\bibinfo  {journal} {Phys. Rev. Lett.}\ }\textbf {\bibinfo {volume} {121}},\ \bibinfo {pages} {218101} (\bibinfo {year} {2018})}\BibitemShut {NoStop}%
\bibitem [{\citenamefont {Nakayama}\ and\ \citenamefont {Toyabe}(2021)}]{naka21}%
  \BibitemOpen
  \bibfield  {author} {\bibinfo {author} {\bibfnamefont {Y.}~\bibnamefont {Nakayama}}\ and\ \bibinfo {author} {\bibfnamefont {S.}~\bibnamefont {Toyabe}},\ }\bibfield  {title} {\bibinfo {title} {Optimal rectification without forward-current suppression by biological molecular motor},\ }\href {https://doi.org/10.1103/PhysRevLett.126.208101} {\bibfield  {journal} {\bibinfo  {journal} {Phys. Rev. Lett.}\ }\textbf {\bibinfo {volume} {126}},\ \bibinfo {pages} {208101} (\bibinfo {year} {2021})}\BibitemShut {NoStop}%
\bibitem [{\citenamefont {Kawai}\ \emph {et~al.}(2007)\citenamefont {Kawai}, \citenamefont {Parrondo},\ and\ \citenamefont {{van den Broeck}}}]{kawa07}%
  \BibitemOpen
  \bibfield  {author} {\bibinfo {author} {\bibfnamefont {R.}~\bibnamefont {Kawai}}, \bibinfo {author} {\bibfnamefont {J.~M.~R.}\ \bibnamefont {Parrondo}},\ and\ \bibinfo {author} {\bibfnamefont {C.}~\bibnamefont {{van den Broeck}}},\ }\bibfield  {title} {\bibinfo {title} {Dissipation: The phase-space perspective},\ }\href {https://doi.org/10.1103/PhysRevLett.98.080602} {\bibfield  {journal} {\bibinfo  {journal} {Phys.\ Rev.\ Lett.}\ }\textbf {\bibinfo {volume} {98}},\ \bibinfo {pages} {080602} (\bibinfo {year} {2007})}\BibitemShut {NoStop}%
\bibitem [{\citenamefont {Gomez-Marin}\ \emph {et~al.}(2008)\citenamefont {Gomez-Marin}, \citenamefont {Parrondo},\ and\ \citenamefont {{van den Broeck}}}]{gome08b}%
  \BibitemOpen
  \bibfield  {author} {\bibinfo {author} {\bibfnamefont {A.}~\bibnamefont {Gomez-Marin}}, \bibinfo {author} {\bibfnamefont {J.}~\bibnamefont {Parrondo}},\ and\ \bibinfo {author} {\bibfnamefont {C.}~\bibnamefont {{van den Broeck}}},\ }\bibfield  {title} {\bibinfo {title} {The "footprints" of irreversibility},\ }\href {https://doi.org/10.1209/0295-5075/82/50002} {\bibfield  {journal} {\bibinfo  {journal} {EPL}\ }\textbf {\bibinfo {volume} {82}},\ \bibinfo {pages} {50002} (\bibinfo {year} {2008})}\BibitemShut {NoStop}%
\bibitem [{\citenamefont {Roldan}\ and\ \citenamefont {Parrondo}(2010)}]{rold10}%
  \BibitemOpen
  \bibfield  {author} {\bibinfo {author} {\bibfnamefont {E.}~\bibnamefont {Roldan}}\ and\ \bibinfo {author} {\bibfnamefont {J.~M.~R.}\ \bibnamefont {Parrondo}},\ }\bibfield  {title} {\bibinfo {title} {Estimating dissipation from single stationary trajectories},\ }\href {https://doi.org/10.1103/PhysRevLett.105.150607} {\bibfield  {journal} {\bibinfo  {journal} {Phys.\ Rev.\ Lett.}\ }\textbf {\bibinfo {volume} {105}},\ \bibinfo {pages} {150607} (\bibinfo {year} {2010})}\BibitemShut {NoStop}%
\bibitem [{\citenamefont {Hartich}\ and\ \citenamefont {Godec}(2024)}]{hart24}%
  \BibitemOpen
  \bibfield  {author} {\bibinfo {author} {\bibfnamefont {D.}~\bibnamefont {Hartich}}\ and\ \bibinfo {author} {\bibfnamefont {A.}~\bibnamefont {Godec}},\ }\bibfield  {title} {\bibinfo {title} {Comment on ``{{Inferring}} broken detailed balance in the absence of observable currents''},\ }\href {https://doi.org/10.1038/s41467-024-52602-0} {\bibfield  {journal} {\bibinfo  {journal} {Nature Communications}\ }\textbf {\bibinfo {volume} {15}},\ \bibinfo {pages} {8678} (\bibinfo {year} {2024})}\BibitemShut {NoStop}%
\bibitem [{\citenamefont {Bisker}\ \emph {et~al.}(2024)\citenamefont {Bisker}, \citenamefont {Mart{\'i}nez}, \citenamefont {Horowitz},\ and\ \citenamefont {Parrondo}}]{bisk24}%
  \BibitemOpen
  \bibfield  {author} {\bibinfo {author} {\bibfnamefont {G.}~\bibnamefont {Bisker}}, \bibinfo {author} {\bibfnamefont {I.~A.}\ \bibnamefont {Mart{\'i}nez}}, \bibinfo {author} {\bibfnamefont {J.~M.}\ \bibnamefont {Horowitz}},\ and\ \bibinfo {author} {\bibfnamefont {J.~M.~R.}\ \bibnamefont {Parrondo}},\ }\bibfield  {title} {\bibinfo {title} {Reply to: {{Comment}} on ``{{Inferring}} broken detailed balance in the absence of observable currents''},\ }\href {https://doi.org/10.1038/s41467-024-52603-z} {\bibfield  {journal} {\bibinfo  {journal} {Nature Communications}\ }\textbf {\bibinfo {volume} {15}},\ \bibinfo {pages} {8679} (\bibinfo {year} {2024})}\BibitemShut {NoStop}%
\bibitem [{Sup()}]{Supplement}%
  \BibitemOpen
  \href@noop {} {}\bibinfo {note} {See Supplemental Material, which contains details about the analytical derivations (Appendix D) and numerical simulations (Appendix E).}\BibitemShut {Stop}%
\bibitem [{\citenamefont {Sekimoto}(2021)}]{seki21}%
  \BibitemOpen
  \bibfield  {author} {\bibinfo {author} {\bibfnamefont {K.}~\bibnamefont {Sekimoto}},\ }\href@noop {} {\bibinfo {title} {Derivation of the first passage time distribution for markovian process on discrete network}} (\bibinfo {year} {2021}),\ \Eprint {https://arxiv.org/abs/arXiv:2110.02216} {arXiv:2110.02216} \BibitemShut {NoStop}%
\end{thebibliography}
%

\pagebreak
\clearpage

\section{End matter}

\paragraph{Appendix A: Response to errors is not bounded by entropy production ---} 

The main text contains an intuitive explanation for why the bounds \eqref{eq:multitr:inequalities} are formulated in terms of the affinity rather than entropy production. This paragraph presents the example in which the ratio $\mathcal{R}^\eps\!/\sigma$ diverges more explicitly. We consider two pairs of transitions, $x$ and $y$, which connect a single state to itself. This peculiarity of having one state and two transition channels is not pivotal but simplifies the discussion; essentially the same reasoning applies to a two-dimensional discrete random walk with horizontal transitions $x$ and vertical transitions $y$. 

We assume that transitions along $y$ are not driven, i.e., $s(y) = 0$, which can be realized through equal transition rates along $y$ and $\rev{y}$. Additionally, we assume that transitions along $x$ and $\rev{x}$ are possible but rare and parametrize $p(x) = e^{s(x)} \delta, p(\rev{x}) = \delta$ for a small $\delta > 0$. The total entropy production in the system is given by
\begin{equation}
    \sigma = (p(x) - p(\rev{x})) s(x) = \delta \cdot s(x) (e^{s(x)} - 1)
,\end{equation}
thus equilibrium is attained for $\delta \to 0$. As error matrix we assume
\begin{equation}
    p_{y' \to x'} = \begin{cases}
        1 \quad y' = y, x' = x \text{ or } y' = \rev{y}, x' = \rev{x} \\
        0 \quad \text{ else }
    \end{cases}
,\end{equation}
which respects the symmetry condition \eqref{eq:symm_error}. The response $\mathcal{R}^\eps$ can be calculated via Eq. \eqref{eq:multitr:generalres}, yielding
\begin{equation}
    \mathcal{R}^\eps = 2 p(y) \left[ \cosh s(x) - 1 \right] \geq p(y) s(x)^2 = \mathcal{R}^\eps_{\text{l.r.}}
\end{equation}
independently of $\delta$. Thus, this quantity remains constant in the limit $\delta \to 0$, whereas $\sigma \to 0$, such that the ratio $\mathcal{R}^\eps\!/\sigma$ diverges.

\paragraph{Appendix B: Transition-based coarse graining---} This section briefly explains estimation of entropy production if some transitions in a network and the time at which they take place are observable, as in the set-up shown in Fig. \ref{fig:err_corr}. 

Sufficient transition statistics enable us to determine the distribution of waiting times between two subsequent transitions $I$ followed by $J$. The corresponding waiting-time distribution $\psi_{I \to J} (t)$ is normalized such that $\sum_J \int_0^\infty dt \psi_{I \to J} (t) = 1$. In the steady state the quantity
\begin{equation}
    \hat{\sigma} = \sum_{IJ} \int_0^\infty dt p(I) \psi_{I \to J} (t) \ln \frac{\psi_{I \to J} (t)}{\psi_{\rev{J} \to \rev{I}} (t)}
    \label{eq:end:sigma_tr}
\end{equation}
satisfies $\hat{\sigma} \leq \sigma$ and is therefore a consistent estimator for entropy production \cite{vdm22,haru22}. The waiting-time distribution $\psi_{I \to J} (t)$ can be viewed as the distribution of first-passage times of $J$, which can be calculated from the dynamics of the underlying microscopic model by solving a suitable initial value problem. For transitions $I = (kl), J = (mn)$ the waiting-time distribution $\psi_{I \to J} (t)$ is given by
\begin{equation}
    \psi_{I \to J} (t) = P(\gamma(t) = m|\gamma(0) = l)k_{mn}
,\end{equation}
where $k_{mn}$ denotes the transition rate from state $m$ to $n$ and $P(\gamma(t) = m|\gamma(0) = l)$ is obtained by solving the master equation subject to the initial condition $P(\gamma(0) = i) = \delta_{il}$ \cite{seki21, vdm22,haru22}. The sequence of observed transitions themselves is itself a discrete-time Markov chain with transition probabilities $\pi_{I \to J}$ given by 
\begin{equation}
     \pi_{I \to J} = \int_0^\infty dt \psi_{I \to J} (t)
.\end{equation}
If removing the observed transitions from the Markov network results in a network that is at detailed balance, e.g., one that does not contain any hidden cycles, then the estimator \eqref{eq:end:sigma_tr} is tight even when waiting times are not taken into account, i.e., the two equalities
\begin{equation}
    \hat{\sigma} = \sigma = \sum_{IJ} p(I) \pi_{I \to J} \ln \frac{\pi_{I \to J}}{\pi_{\rev{J} \to \rev{I}}}
    \label{eq:end:sigma_tr_no_wt}
\end{equation}
are satisfied \cite{vdm22}. The corresponding estimator in the presence of errors is
\begin{equation}
    \hat{\sigma}^\eps = \sum_{IJ} p^\eps(I) \pi^\eps_{I \to J} \ln \frac{\pi^\eps_{I \to J}}{\pi^\eps_{\rev{J} \to \rev{I}}}
    \label{eq:end:sigma_tr_err}
,\end{equation}
with $p^\eps(I)$ as in Eq. \eqref{eq:multitr:ep_def}. As detailed in the SM \cite{Supplement}, the probabilities $\pi^\eps_{I \to J}$ can be calculated from the two-step probabilities $p^\eps(IJ)$ to observe first $I$, then $J$ via
\begin{align}
    \pi^\eps_{I \to J} = \frac{p^\eps(IJ)}{\sum_K p^\eps(IK)} = \frac{p^\eps(IJ)}{p^\eps(I)}
    \label{eq:end:p_split_eps}
.\end{align}
In Figure \ref{fig:err_corr}\,(c), the values for $\sigma$ and $\hat{\sigma}^\eps$ were obtained by first calculating the transition probabilities $\pi_{I \to J}$ from the underlying model followed by a direct calculation using Eq. \eqref{eq:end:sigma_tr_no_wt} and Eq. \eqref{eq:end:sigma_tr_err}, respectively.

It is worth noting that in the error-free case only two of the three pairs of observed transitions are needed to recover the full entropy production rate, i.e., Eq. \eqref{eq:end:sigma_tr_no_wt} can be applied even when only two suitable pairs of transitions are observed, e.g., $K_\pm$ and either $I_\pm$ or $J_\pm$. However, the additional pair of observed transitions becomes crucial when errors are introduced, because it provides the redundant information in the coarse-grained trajectory that enables the error detection and correction mechanism, which leads to the improved estimates of entropy production. 

\paragraph{Appendix C: Decomposition of entropy production---}  
The general result Eq. \eqref{eq:sigma_diff} provides a decomposition of $\hat{\sigma}^\eps$ and $\sigma -\hat{\sigma}^\eps$ into nonnegative contributions from individual pairs of a trajectory $\gamma$ and its observation $\Gamma^\eps$. In the case of observed transitions in a Markov network as in Fig. \ref{fig:err_corr}, the microscopic path weight is known, therefore explicit results like the contribution of parts of a trajectory with $n$ subsequent errors, which scale as $\eps^n$, can be computed explicitly.

The calculation is simplified by utilizing Eq. \eqref{eq:end:sigma_tr_no_wt}, which allows us to, firstly, replace $\gamma$ by its correct coarse graining $\Gamma$ in Eq. \eqref{eq:sigma_diff} and, secondly, discard the waiting times in $\Gamma$ because only the transition statistics of $\Gamma$ contribute to $\sigma$. This leaves us with the task of comparing the irreversibility of $\Gamma^\eps$, the sequence of observed, faulty transitions, to $\Gamma$ the corresponding correct sequence. We now apply the framework of Ref. \cite{vdm23}, which allows a decomposition of the irreversibility of a steady-state trajectory $\Gamma$ of duration $T$ into trajectory snippets $\Gamma^s$ that takes the form
\begin{align}
    \frac{1}{T}\braket{\ln \frac{P[\Gamma]}{P[\widetilde{\Gamma}]}} = \sum_{\Gamma^s, I} P(I) \mathcal{P}[\Gamma^s|I] \ln \frac{\mathcal{P}[\Gamma^s|I]}{\mathcal{P}[\widetilde{\Gamma^s}|\rev{J}]}
    \label{eq:end:snippets}
.\end{align}
A snippet $\Gamma^s$ is defined as a steady-state trajectory between two observable events $I$ and $J$, which satisfy the Markov property in a specific sense. The condition under which Eq. \eqref{eq:end:snippets} applies is that $I$ and $J$ are Markovian events, meaning that, for example, observing $I$ implies that the future dynamics after $I$ is conditionally independent from its past \cite{vdm23}. This condition is satisfied on the level of $\Gamma$ for any observable transitions $I$, $J$, but to obtain the desired decomposition into trajectory snippets containing $n$ subsequent errors only correctly observed transitions are treated as valid initial and final Markovian events. Thus, in the previous equation $P(I) = (1 - \eps) p(I)$ denotes the rate at which a correctly observed transition $I$ occurs and initializes a snippet $\Gamma^s$, which by construction has correctly observed initial and final states but variable length. Schematically we denote $\Gamma^s$ and its observed counterpart, $\Gamma^{s, \eps}$ by
\begin{alignat}{3}
    \Gamma^{s} & = I_0 \to I_1 \to \cdots & & \to I_{n} && \to I_{n+1} \\
    \Gamma^{s, \eps} & = I_0 \to I_1^\eps \to \cdots & & \to I_{n}^\eps &&\to I_{n+1}
\end{alignat}
with respective path weights
\begin{align}
    \mathcal{P}[\Gamma^{s}] & = P(I_0) \mathcal{P}[\Gamma| I_0] \nonumber \\ & = (1 - \eps) p(I_0) \prod_{j = 0}^n \pi_{I_j \to I_{j+1}} \\
    \mathcal{P}[\Gamma^{s, \eps}|\Gamma^{s}] & = \eps^n \prod_{j = 1}^n p_{I_j \to I_j^\eps}
.\end{align}
To apply Eq. \eqref{eq:end:snippets}, we set $I = I_0$, $J = I_{n+1}$ and understand the sum over all possible $\Gamma^s$ as a sum over all possible $I_0, \cdots I_{n+1}$ and all $n \geq 0$. We note that the symmetry condition \eqref{eq:symm_error} is satisfied if $p_{I_j \to I_j^\eps} = p_{\rev{I_j} \to \rev{I_j^\eps}}$. Thus, substituting
\begin{align}
    \ln \frac{\mathcal{P}[\Gamma^s|I_0]}{\mathcal{P}[\widetilde{\Gamma^s}|\rev{I}_{n+1}]} & = \ln \frac{\mathcal{P}[\Gamma^s, \Gamma^{s, \eps}|I_0]}{\mathcal{P}[\widetilde{\Gamma^s}, \widetilde{\Gamma^{s, \eps}}|\rev{I}_{n+1}]} \nonumber \\ & = \ln \frac{\mathcal{P}[\Gamma^{s, \eps}|I_0]}{\mathcal{P}[\widetilde{\Gamma^{s, \eps}}|\rev{I}_{n+1}]} + \ln \frac{\mathcal{P}[\Gamma^s|\Gamma^{s, \eps}]}{\mathcal{P}[\widetilde{\Gamma^{s}}|\widetilde{\Gamma^{s, \eps}}]}
\end{align}
into \eqref{eq:end:snippets} yields the decomposition into
\begin{equation}
    \hat{\sigma}^\eps = \sum_{n \geq 0} \sum_{\substack{I_0, ..., I_{n+1} \\ I^\eps_1, ..., I^\eps_n}} \mkern-12mu P(I_0) \mathcal{P}[\Gamma^{s, \eps}, \Gamma^{s}|I_0] \ln \frac{\mathcal{P}[\Gamma^{s, \eps}|I_0]}{\mathcal{P}[\widetilde{\Gamma^{s, \eps}}|\rev{I}_{n+1}]}
    \label{eq:end:decomposition}
,\end{equation}
the visible contribution under errors, and its complement $\sigma - \hat{\sigma}^\eps$ in analogy to Eq. \eqref{eq:sigma_diff}. We can now identify $\hat{\sigma}^\eps_n$ as defined in Eq. \eqref{eq:error:decomposition} and $\sigma_n - \hat{\sigma}^\eps_n$ as the contributions due to the $n$-th term, which contain all possible snippets that contain $n$ subsequent errors surrounded by a correctly registered initial and final transition.

\widetext

\appendix

\clearpage

\section{Supplemental Material for "Thermodynamic bounds and error correction for faulty coarse graining"}

\subsection{Appendix D: Thermodynamic bounds -- Calculations and proofs}

The following sections contain details about the derivations of the exact expression Eq. \eqref{eq:multitr:generalres} for the response quantity $\mathcal{R}^\eps$ defined in Eq. \eqref{eq:multitr:defR}. 

\subsubsection{First-order response in terms of probabilities}

In this section we calculate $\mathcal{R}^\eps$ in terms of $p(x)$, the steady-state rates of a transition $x$, and $p_{y \to x}$, the probability that an error in transition $y$ results in a registered transition $x$. In the following calculations, we use the shorthand $q(x) = p(\rev{x})$ and similarly $q^\eps(x) = p^\eps(\rev{x})$ to denote the steady-state rate of the reverse transition of $x$.

Starting from the expression
\begin{align}
    \braket{\sigma} - \braket{\sigma^\eps} = \frac{1}{2} \left( \sum_x (p(x) - q(x)) \ln \frac{p(x)}{q(x)} - \sum_x  (p^\eps(x) - q^\eps(x)) \ln \frac{p^\eps(x)}{q^\eps(x)} \right)
,\end{align}
we first express $p^\eps(x)$ in terms of $p(x)$ using Eq. \eqref{eq:multitr:ep_def}. We note that the following calculations remain valid for a nonvanishing diagonal $p_{y \to y} \geq 0$. Expanding the logarithm to second order in $\eps$ yields
\begin{align}
    \ln \frac{p^\eps(x)}{q^\eps(x)} & = \ln \frac{p(x) (1 - \eps) + \sum_y  \eps p(y) p_{y \to x}}{q(x) (1 - \eps) + \sum_y  \eps q(y) p_{y \to x}} 
    = \ln \frac{p(x)}{q(x)} + \ln \frac{1 - \eps + \eps \sum_y p(y) p_{y \to x}/p(x)}{1 - \eps + \eps \sum_y q(y) p_{y \to x}/q(x)} \\
    & = \ln \frac{p(x)}{q(x)} + \eps \left( \frac{\sum_y p(y) p_{y \to x}}{p(x)} - \frac{\sum_y q(y) p_{y \to x}}{q(x)} \right) + O(\eps^2)
,\end{align}
which can be substituted into the previous equation to obtain 
\begin{align}
    2 (\braket{\sigma} - \braket{\sigma^\eps}) = \, & \sum_x \left( p(x) - q(x) - (p^\eps(x) - q^\eps(x)) \right) \ln \frac{p(x)}{q(x)} \nonumber \\ & - \sum_x (p^\eps(x) - q^\eps(x)) \left( \eps \left( \frac{\sum_y p(y) p_{y \to x}}{p(x)} - \frac{\sum_y q(y) p_{y \to x}}{q(x)} \right) + O(\eps^2) \right)
.\end{align}
For the first line, we use Eq. \eqref{eq:multitr:ep_def} in the form
\begin{align}
    (p(x) - q(x)) - (p^\eps(x) - q^\eps(x)) = - \eps \left( - p(x) + q(x) + \sum _y p(y) p_{y \to x} - \sum _y q(y) p_{y \to x} \right) + O(\eps^2)
,\end{align}
whereas the second line already is of order $\eps$, so the zeroth-order approximation of $p^\eps(x) - q^\eps(x)$ suffices. After rearranging terms, we take the limit $\eps \to 0$
\begin{align}
    2 \mathcal{R}^\eps = 2 \lim_{\eps \to 0} \frac{\braket{\sigma} - \braket{\sigma^\eps}}{\eps} 
    = & \, \sum_{x} (p(x) - q(x)) \ln \frac{p(x)}{q(x)} - \sum_{xy} p_{y \to x} (p(y) - q(y)) \ln \frac{p(x)}{q(x)} \nonumber \\
    & - \sum_x (p(x) - q(x)) \left( \frac{\sum_y p(y) p_{y \to x}}{p(x)} - \frac{\sum_y q(y) p_{y \to x}}{q(x)}  \right)
    \label{eq:app_a:general_res}
.\end{align}
The first term on the right hand side is $2 \braket{\sigma}$, but it will be convenient to artificially introduce a normalization $\sum_x p_{y \to x}=1$ to rewrite this term as $2 \braket{\sigma} = \sum_{yx} p_{y \to x}(p(y)-q(y) \ln [p(y)/q(y)]$. After dividing by $2$ and rewriting the third term on the right hand side, we arrive at
\begin{align}
    \mathcal{R}^\eps
    = \, \frac{1}{2} \left( \sum_{yx} p_{y \to x}(p(y)-q(y) \ln \frac{p(y)}{q(y)} - \sum_{xy} p_{y \to x} (p(y) - q(y)) \ln \frac{p(x)}{q(x)}
    - 2 +\sum_{xy} p_{y \to x} \left( \frac{q(x)}{p(x)} p(y) + \frac{p(x)}{q(x)} q(y) \right) \right)
    \label{eq:app_a:general_res_2}.
\end{align}
A more symmetric form can be obtained by combining the first two logarithmic terms, which yields
\begin{align}
    \mathcal{R}^\eps
    & = -1 + \frac{1}{2} \sum_{x,y \neq x} p_{y \to x} (p(y) - q(y)) \ln \frac{p(y)q(x)}{q(y)p(x)} + \frac{1}{2} \sum_{xy} p_{y \to x} \left( \frac{q(x)}{p(x)} p(y) + \frac{p(x)}{q(x)} q(y) \right) \nonumber \\
    & = -1 + \frac{1}{2} \sum_{xy} p_{y \to x} p(y) \left[\frac{q(y)p(x)}{p(y)q(x)} + \ln \frac{p(y)q(x)}{q(y)p(x)} \right] + \frac{1}{2} \sum_{xy} p_{y \to x} q(y) \left[ \frac{p(y)q(x)}{q(y)p(x)} + \ln \frac{q(y)p(x)}{p(y)q(x)} \right]
    \label{eq:app_a:general_res_3} 
.\end{align}

\subsubsection{Response in terms of affinities}

The previous expression, Eq. \eqref{eq:app_a:general_res_3} contains two sums, into which we distribute the constant $-1$ symmetrically. After identifying $s(x) = \ln [p(x)/q(x)]$ and rearranging, we obtain
\begin{align}
\mathcal{R}^\eps
     & = \frac{1}{2} \sum_{x,y \neq x} p_{y \to x} p(y) \left[e^{s(x) - s(y)} - (s(x) - s(y)) - 1 \right] + \frac{1}{2} \sum_{xy} p_{y \to x} q(y) \left[e^{s(y) - s(x)} - (s(y) - s(x)) - 1 \right] \nonumber \\
     & = \sum_{x,y \neq x} p_{y \to x} p(y) \left[e^{s(x) - s(y)} - (s(x) - s(y)) - 1 \right]
     \label{eq:app_a:resp_s}
,\end{align}
which corresponds to Eq. \eqref{eq:multitr:generalres} in the main text. In passing to the second line in Eq. \eqref{eq:app_a:resp_s}, we have replaced the summation indices $x,y$ by $\tilde{x}, \tilde{y}$ in the second sum. Using the symmetry condition $p_{y \to x} = p_{\tilde{y} \to \tilde{x}}$ and the antisymmetry $s(\tilde{x}) = - s(x)$,  we see that the two terms in Eq. \eqref{eq:app_a:resp_s} are identical, i.e.,
\begin{align}
    \frac{1}{2} \sum_{xy} p_{y \to x} q(y) \left[e^{s(y) - s(x)} - (s(y) - s(x)) - 1 \right] & = \frac{1}{2} \sum_{xy} p_{y \to x} q(\tilde{y}) \left[e^{s(\tilde{y}) - s(\tilde{x})} - (s(\tilde{y}) - s(\tilde{x})) - 1 \right] \nonumber \\
    & =  \frac{1}{2} \sum_{xy} p_{y \to x} p(y) \left[e^{s(x) - s(y)} - (s(x) - s(y)) - 1 \right]
.\end{align}

\subsubsection{Proof of the lower bound}

Starting from the result \eqref{eq:app_a:resp_s}, it suffices to use the inequality $e^z \geq z + 1$ to confirm nonnegativity of $\mathcal{R}^\eps$. To get a stronger lower bound in terms of
\begin{equation}
\Delta s_\text{min} = \min_{x \neq y} |s(x) - s(y)|
,\end{equation}
we include the second-order term in the expansion $e^z - z -1 \geq z^2/2$, valid for real $z$, which yields
\begin{align}
\mathcal{R}^\eps 
     & = \sum_{x,y \neq x} p_{y \to x} p(y) \left[e^{s(x) - s(y)} - (s(x) - s(y)) - 1 \right]
     \geq \frac{1}{2} \sum_{x,y \neq x} p_{y \to x} p(y) \left( s(x) - s(y) \right)^2 \nonumber \\
     & \geq \frac{1}{2} (1 - \sum_y p(y) p_{y \to y}) \Delta s_\text{min}^2
.\end{align}
These results correspond to the respective bounds involving $\mathcal{R}^\eps_{\text{l.r.}}$ and $\mathcal{R}^\eps_{\text{min}} (\Delta s_\text{min})$ in the inequalities \eqref{eq:multitr:inequalities} and \eqref{eq:num:lb}.

\subsubsection{Proof of the upper bound}

Again, we start from Eq. \eqref{eq:app_a:resp_s}, which can be rewritten as
\begin{align}
    \mathcal{R}^\eps = \frac{1}{2} \sum_{x,y \neq x} p_{y \to x} \left[ \left(p(y) - p(\rev{y}) \right) \left(s(y) - s(x) \right) + \left( - p(y) - p(\rev{y}) + p(y) e^{s(x) - s(y)} + p(\rev{y}) e^{s(y) - s(x)}  \right) \right]
.\end{align}
We discuss the two terms in the square brackets separately. First, since $s(y) = \ln[p(y)/p(\rev{y})]$, we have
\begin{equation}
    p(y) - p(\rev{y}) = \left(p(y) + p(\rev{y}) \right) \frac{e^{s(y)} - 1}{e^{s(y)} + 1} = \left(p(y) + p(\rev{y}) \right) \tanh \left( \frac{s(y)}{2} \right)
    \label{eq:app_a:upper:aux0}
.\end{equation}
This identity can be used to rewrite the term
\begin{align}
    \frac{1}{2} \sum_{x,y \neq x} p_{y \to x} (p(y) - p(\rev{y}))(s(y) - s(x)) & = \frac{1}{2} \sum_{x,y \neq x} \left(p(y) + p(\rev{y}) \right) \tanh \left( \frac{s(y)}{2} \right) (s(y) - s(x))  \nonumber \\
    & \leq \frac{1}{2} \sum_{x,y \neq x} p_{y \to x} \left(p(y) + p(\rev{y}) \right) \tanh \left( \frac{|s(y)|}{2} \right) |s(y) - s(x)|  \nonumber \\
    & \leq \frac{1}{2} \sum_{x,y \neq x} p_{y \to x} \left(p(y) + p(\rev{y}) \right) \tanh \left( \frac{\smax}{4} \right) \smax  \nonumber \\
    & = 2 \left( \tanh \frac{\smax}{4} \right) \frac{\smax}{2} 
    \label{eq:app_a:upper:aux2}
.\end{align}
The first inequality involves taking the absolute value of each term, whereas the second one makes use of $\smax = \max_{x,y} |s(x) - s(y)| = 2 \max_{x} |s(x)|$. In passing to the final line, we use normalization in the form $\sum_{x,y \neq x} p_{y \to x} p(y) = \sum_{x,y \neq x} p_{y \to x} p(\rev{y}) = 1$.

Let us now discuss the second term, 
\begin{equation}
    R := \frac{1}{2} \sum_{x,y \neq x} p_{y \to x} \left[ - p(y) - p(\rev{y}) + p(y) e^{s(x) - s(y)} + p(\rev{y}) e^{s(y) - s(x)}  \right]
    \label{eq:app_a:upper:aux1}
.\end{equation}
By using $p(y) e^{- s(y)} = p(\rev{y})$ and $p(\rev{y}) e^{s(y)} = p(y)$, the expression inside the square brackets can be rewritten as
\begin{align}
    - p(y) - p(\rev{y}) + & \, p(y) e^{s(x) - s(y)} + p(\rev{y}) e^{s(y) - s(x)} = 2 \sinh \frac{s(x)}{2} \left[  p(\rev{y}) e^{\frac{s(x)}{2}} - p(y) e^{- \frac{s(x)}{2}} \right] \nonumber \\
    & = 2 \sinh \frac{s(x)}{2} \left[ (p(\rev{y}) - p(y)) \cosh \frac{s(x)}{2} + (p(\rev{y}) + p(y)) \sinh \frac{s(x)}{2} \right] \nonumber \\
    & = 2 \sinh \frac{s(x)}{2} \cosh \frac{s(x)}{2} (p(\rev{y}) - p(y))  + 2 \sinh^2 \frac{s(x)}{2}  (p(\rev{y}) + p(y))
.\end{align}
We now make use of the identity $2 \sinh \frac{t}{2} \cosh \frac{t}{2} = \sinh t$ and its reformulation $2 \sinh^2 \frac{t}{2} = \tanh \frac{t}{2} \sinh t$ for $t = s(x)$. Substituting the result into Eq. \eqref{eq:app_a:upper:aux1}, we obtain
\begin{align}
    R = & \, \frac{1}{2} \sum_{x,y \neq x} p_{y \to x} \left[ (p(\rev{y}) - p(y)) \sinh s(x) + (p(\rev{y}) + p(y)) \tanh \frac{s(x)}{2} \sinh s(x) \right] \nonumber \\
    = & \, \frac{1}{2} \sum_{x,y \neq x} p_{y \to x} (p(\rev{y}) + p(y)) \sinh s(x) \left[- \tanh \frac{s(y)}{2} + \tanh \frac{s(x)}{2}  \right]
\end{align}
after using Eq. \eqref{eq:app_a:upper:aux0}. This expression can be bounded by its absolute value, which results in the bound
\begin{align}
    R \leq \frac{1}{2} \sum_{x,y \neq x} p_{y \to x} (p(\rev{y}) + p(y)) \sinh |s(x)| \left[\tanh \frac{|s(y)|}{2} + \tanh \frac{|s(x)|}{2}  \right]
    \leq 2 \sinh \frac{\smax}{2} \tanh \frac{\smax}{4}
\end{align}
after identifying the upper bound in terms of $\smax = \max_{x,y} |s(x) - s(y)| = 2 \max_{x} |s(x)|$ and using normalization $\sum_{x,y \neq x} p_{y \to x} p(y) = \sum_{x,y \neq x} p_{y \to x} p(\rev{y}) = 1$. Combining this result and Eq. \eqref{eq:app_a:upper:aux2}, we obtain the upper bound
\begin{align}
    \mathcal{R}^\eps \leq \mathcal{R}^\eps_{\text{max}} (\Delta s_\text{max}) = 2 \tanh \frac{\smax}{4} \left( \frac{\smax}{2} + \sinh \frac{\smax}{2} \right)
\end{align}
stated as Eq. \eqref{eq:multitr:inequalities} in the main text. 

In the case of the one-dimensional asymmetric random walk, there are only two possible transitions, i.e., $x,y = \pm$. Additionally, there are no degrees of freedom in $p_{x\to y}$, so that
\begin{align}
    p_{y\to x} = \begin{cases} 1 \quad & x=+, y=- \text{ or } y=+, x=- \\ 0 \quad & \text{ else} \end{cases}
.\end{align}
Additionally, $\ln [p(+)/p(-)] = s = \max_{x,y} |s(x) - s(y)|/2 = \Delta s_\text{max}/2$ and its negative are the only nontrivial affinities, so that one can confirm that the inequalities in the derivation of $\mathcal{R}^\eps \leq \mathcal{R}^\eps_{\text{max}} (\Delta s_\text{max})$ become equalities. Thus, we obtain
\begin{align}
    \mathcal{R}^\eps = 2 \tanh \frac{s}{2} \left( s + \sinh s \right)
\end{align}
for the asymmetric random walk.

\subsection{Appendix E: Simulation parameters}

This section presents the model parameters used in the simulations that generate the data shown in Figures 1 and 2 in the main paper.

\subsubsection{Figure 1}

For a given number $n = 1, 2, 3, 10$ of transition pairs, we generate $N = 5 \cdot 10^4$ systems with $p(x) = u(x)/\sum_x u(x)$ and i.i.d. $u(x)$ chosen from a uniform distribution $U(0,1)$. The error matrix is created according to $p_{y \to x} = u(y,x)/(\sum_x u(y,x))$ to ensure normalization. The $u(y,x)$ satisfy $u(y,x) = 0$ if $x = y$, $u(y,x) = u(\rev{y},\rev{x})$ to respect the symmetry condition (main text, Eq. (3)) and are otherwise i.i.d. and chosen from a uniform distribution $U(0,1)$. 

\subsubsection{Figure 2}

The microscopic model has the topology depicted in Fig. 2 (a). Transition rates $k_{ij}$ from state $i$ to state $j$ are given as $k_{31} = k_{12} = k_{24} = 2$, $k_{13} = k_{21} = k_{42} = 3$, $k_{46} = k_{64} = k_{65} = k_{56} = k_{53} = k_{35} = 0.4$. A variable affinity $f$ (in units of energy per $k_BT$) is applied at the transition $3 \to 4$, which must be incorporated into the model such that $\ln (k_{34}/k_{43}) = f$. In this model, we parametrize the transition rates as $k_{34}(f) = 3 \exp(f/2)$ and $k_{43}(f) = 2 \exp(-f/2)$. The error matrix $p_y \to x$ is given by
\begin{equation}
    p_{y \to x} = \begin{cases}
        0.6 \quad & x = \rev{y} \\
        0 \quad & x = y \\
        0.1 \quad & \text{else} \\
    \end{cases}
,\end{equation}
i.e., an erroneously detected transition has a higher probability to be detected as its reverse than to be detected as any of the other ones. The estimator for entropy production, 
\begin{equation}
    \hat{\sigma}^\eps = \sum_{IJ} p^\eps(I) \pi^\eps_{I \to J} \ln \frac{\pi^\eps_{I \to J}}{\pi^\eps_{\rev{J} \to \rev{I}}}
,\end{equation}
is evaluated as discussed in Appendix A in the End Matter, with the explicit expression for $\pi^\eps_{I \to J} = p^\eps(IJ)/p^\eps(I)$ (cf. Eq. \eqref{eq:end:p_split_eps}) given by
\begin{align}
    \pi^\eps_{I \to J} = \frac{(1 - \eps)^2 p(I) \pi_{I \to J} + \eps (1-\eps) \sum_X \left(p(I) \pi_{I \to X} p_{X \to J} + p(X) \pi_{X \to J} p_{X \to I}  \right) + \eps^2 \sum_{X,Y} p(X) \pi_{X \to Y} p_{X \to I} p_{Y \to J}}{(1 - \eps) p(I) + \eps \sum_X p(X) p_{X \to I}}
.\end{align}
The summation over the individual paths was performed according to Eq. \eqref{eq:end:decomposition} was performed up to $n = 5$, i.e., includes snippets with up to $5$ consecutive errors. Adding all contributions together, we numerically confirm that $\sum_{n = 0}^5 \sigma_n$ deviates from the true value $\sigma$ by less than $0.5\%$ relative error.

\end{document}